\documentclass[review]{elsarticle}
\usepackage{graphicx}
\usepackage{float}
\usepackage{mathtools}
\usepackage{color}
\usepackage{algorithm}
\usepackage{algorithmic}
\usepackage{epsfig}
\usepackage{amssymb}
\usepackage{bm}
\usepackage{bbm}

\journal{Journal of \LaTeX\ Templates}

\begin{document}

\begin{frontmatter}
\title{Confidence Intervals and Hypothesis Testing for the Permutation Entropy with an application to Epilepsy}

\author[itba,unla]{Francisco Traversaro\corref{cor1}}
\ead{ftraversaro@itba.edu.ar}
\author[hi]{Francisco Redelico}
\ead{franciscoredelico@gmail.com}

\cortext[cor1]{Corresponding author}
\address[itba]{Instituto Tecn\'olgico de Buenos Aires, 
               Av.~Eduardo Madero 399 - (C1181ACH) Ciudad  Aut\'onoma de Buenos Aires, Argentina}
\address[unla]{CONICET - Universidad Nacional de Lan\'us, Grupo de Investigaci\'on en sistemas de informaci\'on,
           29 de Septiembre 3901, B1826GLC Lan\'us, Buenos Aires.}
\address[hi]{ CONICET - Hospital Italiano de Buenos Aires, Departamento de Inform\'atica en Salud, 
              Per\'on 4190 - (C1199ABB) Ciudad  Aut\'onoma de Buenos Aires, Argentina.}

\begin{abstract}
In nonlinear dynamics, and to a lesser extent in other fields, a widely used measure of complexity is the Permutation Entropy.  But there is still no known method to determine the accuracy of this measure. There has been little research on the statistical properties of this quantity that characterize time series. The literature describes some resampling methods of quantities used in nonlinear dynamics - as the largest Lyapunov exponent - but all of these seems to fail. In this contribution we propose a parametric bootstrap methodology using a symbolic representation of the time series in order to obtain the distribution of the Permutation Entropy estimator. We perform several time series simulations given by well known stochastic processes: the $1/f^{\alpha}$ noise family, and show in each case that the proposed accuracy measure is as efficient as the one obtained by the frequentist approach of repeating the experiment. The complexity of brain electrical activity, measured by the Permutation Entropy, has been extensively used in epilepsy research for detection in dynamical changes in electroencephalogram (EEG) signal with no consideration of the variability of this complexity measure. An application of the parametric bootstrap methodology is used to compare normal and pre-ictal EEG signals. 
\end{abstract}


\end{frontmatter}


\section{Introduction}
\label{sec:intro}
In 2002, Bandt and Pompe introduced a measure of complexity for time series \citep{bandt2002permutation} named \textit{permutation entropy} (PE). It is an information entropy \citep{gray2011entropy} that takes account of the time evolution of the time series, in contrast with other prominent information entropies as the Shannon entropy \citep{shannon2001mathematical}. It computation is fast, requires not too long time series \citep{riedl2013practical} and it is robust against noise \citep{quintero2015numerical}. This measure has been widely used in non-linear dynamics \citep{keller2005ordinal,de2008randomizing,rosso2010generalized,masoller2011quantifying}, and to a lesser extent in Stochastic Processes \citep{rosso2007extracting,sinn2011estimation,zunino2008permutation}, among others. It has also had a great impact in in such different and important areas of applied science and engineering as varied as Mechanics Engineering \citep{yan2012permutation,redelico2017evaluation}, Epilepsy \citep{redelico2017classification,olofsen2008permutation}, anaesthesia \citep{jordan2008electroencephalographic}, Cardiology \citep{frank2006permutation,parlitz2012classifying}, Finance \citep{matilla2009detection}, Climate Change \citep{carpi2013analysis}.\\
Since its publication and up to the end of 2016, this paper has been cited in 789 papers, according Scopus bibliographic database, and the evolution of the cites seems to indicate that it will be increasing within time. All these facts made an investigation of PE from the statistic point of view an important issue.\\

There has been little research, up to our knowledge, on the statistical properties of the quantities used in nonlinear dynamics to characterize time series. This lack of research may be due the lack of distributional theory of these quantitites,  yielding resampling technics as the most powerfull tool to overcome this task. Perhaps one exception to this is the research on the distribution of the largest Lyapunov exponent and the correlation dimension\citep{boeing2016visual}. We will summarize one of the most important discussion in this matter, according to our criterion and having in mind the computational scheme that we are proposing: In \citep{genccay1996statistical} a methodology to calculate the empirical distributions of Lyapunov exponents based on a a traditional bootstrapping technique is presented, providing a formal test of chaos under the null hypothesis. However, in \citep{ziehmann1999bootstrap} it is shown that the previously bootstrap approach seems to fail to provide reliable bounds for estimates the Lyapunvos exponents, and conclude that the traditional bootstrap cannot be applied for estimating multiplicative ergodic statistics. In \citep{brzozowska2004application}, a moving blocks bootstrap procedure is used to detect a positive Lyapunov exponent in financial time series. However, the time series generated by moving block bootstrap present artifacts which are caused by joining randomly selected blocks, so the serial dependence is preserved within, but not between, the block.\\

%
%

Regarding time series symbolic dynamics, in \citep{bandt2007order} the probabilities generated using the Bandt and Pompe methodology are calculated analytically for Gaussian Processes for symbol length equal three, but they recognize that for larger length this is not possible, for that reason a computer based method is required to estimate the bias and variance in the PE estimation.\\

In this contribution we propose a different simulation method (i.e parametric bootstrap) for estimating the bias, variance and confidence intervals for the Permutation Entropy estimation, along with hypothesis testing, that consists in simulate bootstrap symbolic time series samples that are thought to be produced by a probabilistic model with a fixed transition probability extracted from the original time series.

%
%
 
In order to show some results from our method we simulate a well known family of time series: the $1/f^{\alpha}$ noise. We compute bias, variance and confidence intervals for the Permutation Entropy of these time series according time series length and several parameters. In addition, an application of the parametric bootstrap methodology for hypothesis testing is used to compare normal and pre-ictal EEG signals. 


The paper reads as follows: Section \ref{sec:permu} shows a brief review of PE in order to present the estimator to be evaluated using the bootstrap approach, Section \ref{Bootgeneral} presents and explains the proposed parametric bootstrap, firstly a brief review of the bootstrap scheme is done as introduction to our method, then in Subsection \ref{transition} the core of the bootstrap approach is presented, i.e. the probability transitions computation is explaiedn and finally in Subsection \ref{boot2} the algortihm to parametric bootstrap PE is explained. Section \ref{numerical} presents the dynamical systems simulated, Section \ref{EEGdata} introduces the experimental data used in the application and Section \ref{results} is devoted to the results and conclusions of this contribution. 


\section{Permutation Entropy}
\label{sec:permu}
In this Section we briefly review the PE to make the article self contained and accessible for a wider audience.\\

\begin{figure}[h!]
\includegraphics[width=\textwidth]{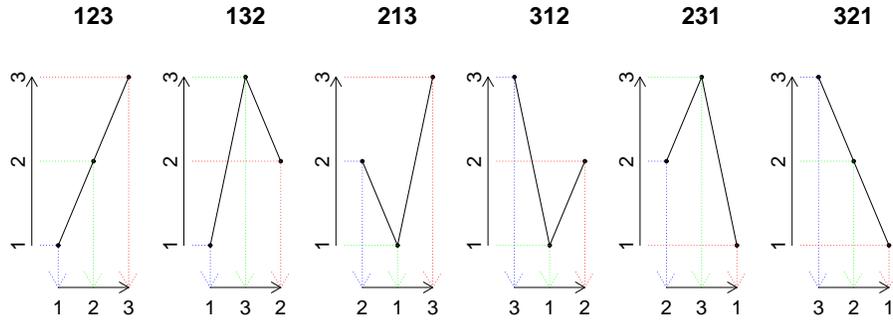}
\caption{\textbf{Rank Permutation Mapping} All symbols for $m=3$ are shown. With this Rank Alphabet one simply maps each value $x_i$ in $X_m(t)$ placing its rank $R(x_i) \in \{1,2,\dots,m\}$ in chronological order to form $\pi_i$ in $S_m$.It can be seen that the indexes of the vertical axis are fixed, ordered by amplitude (i.e ranks), and they are mapped onto the time axis. For each pattern $X_3(t)=(x_t,x_{t+1},x_{t+2})$, the resultant symbol can be obtained reading the labels in the horizontal axis from left to right (in chronological order). }
\label{fig:rank}
\end{figure}

\begin{figure}[h!]
\includegraphics[width=\textwidth]{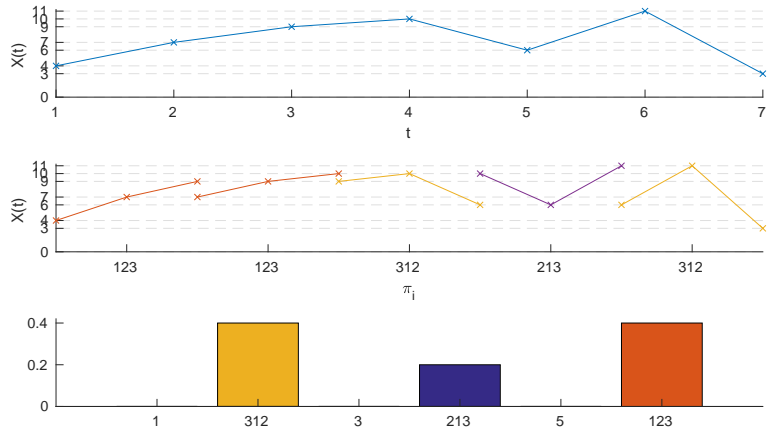}
\caption{Example of the calculation of the permutation entropy.
(top) Time series $X_t=(4,7,9,10,6,11,3)$. (middle) symbols $\pi_i$ generated from the time series using the Rank Permutation Map. (bottom) relative frecuency of $S_{m=3}$ elements for the exemplified time series, $P(312)=0.4$, $P(123)=0.4$ and $P(213)=0.2$.}
\label{fig:ejemplo}
\end{figure}

Let $\{X_t\}_{t \in T}$ be a realization of a data generator process in form of a real valued time series of length $T \in \mathbb{N}$. A measure of uncertaintly about $\{X_t\}_{t \in T}$ is the \textit{normalized} Shannon entropy \citep{shannon2001mathematical} ($0 \leq {\mathcal H} \leq 1$), which is defined as:
\begin{equation}
\label{shannon-disc}
{\mathcal H}[P]~=~ S[P]  / S_{max} ~=~\left\{-\sum_{i=1}^{N}~P_i~\ln( P_i) \right\} /  S_{max} \ ,
\end{equation}
where $P_i$ is a probability to be extracted from the time series, $N$ is the cardinality of the $P_i$ set $\{p_i\}_1^N$, the denominator  $S_{max} = S[P_e] = \ln N$ is obtained by a uniform probability distribution $P_e = \{P_i =1/N,~ \forall i = 1, \cdots, N\}$.\\
Bandt and Pompe proposed a symbolization technique to estimate $P_i$ and compute PE, $\hat{\mathcal H}(m,\tau)$. First we recall that PE has two tuning parameters, i.e. $m$ the symbol length and $\tau$ the time delay. Within this paper, we set $\tau=1$ with no loss of generality and it will omitted, so we will use $\mathcal H = \mathcal H(m)$ for sake of simplicity. Let $X_m(t)=(x_t,x_{t+1},\dots,x_{t+m-1})$ with $0\leq t \leq T-m+1$ be a non-disjoint partition containing the vectors of real values of length $m$ of the time series $\{X_t\}_{t \in T}$. Let $S_{m \geq 3}$ the symmetric group of order $m!$ form by all possible permutation of order $m$, $\pi_i=(i_1,i_2,\dots,i_m) \in~S_m$ ($i_j \neq i_k \forall j \neq k$ so every element in $\pi_i$ is unique). We will call an element $\pi_i$ in $S_m$ a symbol or a motiv as well. Then $X_m(t)$ can be mapped to a symbol $\pi_i$ in $S_m$ for a given but otherwise arbitrary $t$. The $m$ number of real values $X_m(t)=(x_t,x_{t+1},\dots,x_{t+m-1})$ are mapped onto their rank. The rank function is defined as: 
\begin{equation} 
R(x_{t+n}) = \sum_{k=0}^{m-1} \mathbbm{1}(x_{t+k} \leq x_{t+n})
\end{equation}
where $ \mathbbm{1}$ is the indicator function (i.e $ \mathbbm{1}(Z) = 1$ if $Z$ is true and $0$ otherwise) , $x_{t+n} \in X_m(t)$ with $0\leq n \leq m-1$ and $1 \leq R(x_{t+n}) \leq m$. So the rank $R(min(x_{t+k}))=1$ and $R(max(x_{t+k}))=m$. The complete alphabet is all the possible permutation of the ranks.   Hence, any vector $X_m(t)$ is uniquely mapped onto $\pi_i=\left(R(x_t),R(x_{t+1}),\dots,R(x_{t+m-1})\right) \in S_m$. With this  Rank Permutation Mapping one simply maps each value $x_i$ in $X_m(t)$ placing its rank $R(x_i) \in \{1,2,\dots,m\}$ in chronological order to form $\pi_i$ in $S_m$.  In Figure \ref{fig:rank} an illustrative drawing of this mapping for all alternatives in $m=3$ is presented. It can be seen that the indexes of the vertical axis are fixed, ordered by amplitude (i.e ranks), and they are mapped onto the time axis. The resultant symbol can be obtained reading the labels in the horizontal axis from left to right (in chronological order). This method is used by \citep{autocorrelationBandt,riedl2013practical,Bandt2005} among others.
For example, let us take the series with seven values ($T=7$) \citep{bandt2002permutation} (see Fig. \ref{fig:ejemplo}, top), and motiv length $m=3$:
\begin{equation}
X_t=(4,7,9,10,6,11,3)
\end{equation}
$X_3(1)=(4,7,9)$ and $X_3(2)=(7,9,10)$ represents the permutation $\pi = 123$ since $R(x_1)= 1$ ,$R(x_2)= 2$, $R(x_3)= 3$. $X_3(3)=(9,10,6)$ and $X_3(4)=(6,11,3)$ correspond to the permutation $\pi = 231$ since $R(x_1)= 2$ ,$R(x_2)= 3$, $R(x_3)= 1$ (see Fig. \ref{fig:ejemplo}, middle). Using the rank permutation Mapping we compute $P(\pi_i)$ (see Fig. \ref{fig:ejemplo}, bottom) ,
\begin{equation}
 P(\pi_i)=\frac{\sum_{l=1}^{T-m+1}\mathbbm{1}(X_m(l) \text{ has ordinal patter } \pi_i \text{ in } S_m)}{T-m+1},
\end{equation}
where $ \mathbbm{1}$ is the indicator function  and $i=1,\dots,m!$. Using these probabilities, $\hat{\mathcal H}(m)$ can be computed as,
\begin{equation}
\label{eq:shannon}
\hat{\mathcal H}(m)=\left\{-\sum_{i=1}^{N}~P(\pi_i)~\ln( P(\pi_i)) \right\} /  S_{max} \ ,
\end{equation}
 where $N=m!$ is the order of the symmetric group $S_m$ and $S_{max}=log(N)$.\\

\section{The Bootstrap approach}
\label{Bootgeneral}
The bootstrap is a computer based method for assigning measures of accuracy to the desired statistical variable estimates. If $\mathcal{H}$ is an unknown characteristic of a model $\Psi$, an estimator $\hat{\mathcal{H}}$ can be derived from the sample generated by $\Psi$ in a single experiment. A way to obtain the distribution of $\hat{\mathcal{H}}$ is to repeat the experiment a large number of times and approximate the distribution of $\hat{\mathcal{H}}$ by the so obtained empirical distribution. In most practical situations this method is impossible because the experiment is not reproducible, or is unenforceable for cost reasons. 
The spirit of the  bootstrap methodology is to estimate the sampling distribution of a statistic (i.e  quantifier or a parameter estimator) from the data at hand by analogy to the 'thought experiment" that motivates the sampling distribution.

\begin{figure}
\includegraphics[width=\textwidth]{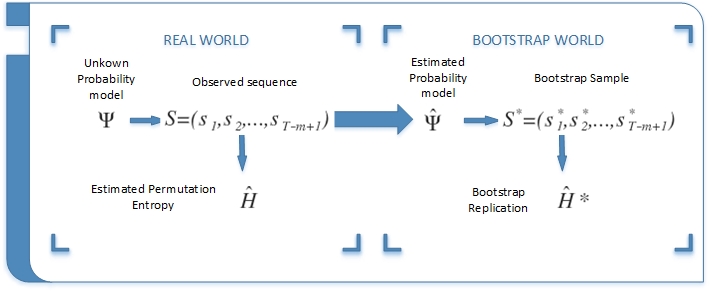}
\caption{Schematic diagram of the parametric bootstrap approach. An unknown probability model, $\Psi~=~\Psi(P^{ij})$, gives an observed sequence $S$ from which we estimate $\hat{{\mathcal H}}$ , so the bootstrap approach suggest to estimate this model $\hat{\Psi}~=~\Psi(\hat{P}^{ij})$ and get the correspondent bootstrap samples $S^{*}$ from which we estimate $\hat{{\mathcal H}}^{*}$}
\label{fig:bootscheme}
\end{figure}

Suppose that an unknown probability model $\Psi$ gives an observed data set $\textbf{X} = \{x_t\}_{t=1}^n$ by random sampling and let $\hat{\theta}_{\Psi}(\textbf{X},T)$ be the statistic of interest that estimates our true value $\theta = f(\Psi)$. Then with the observed data set $\textbf{X}$ we produce an estimate $\hat{\Psi}$. The trick is now to repeat the random sampling but with the estimate $\hat{\Psi}$ giving bootstrap samples $\textbf{X}^{*} = \{x^{*}_t\}_{t=1}^n$ and for each bootstrap sample we calculate $\hat{\theta}^{*}_{\hat{\Psi}}(\textbf{X}^{*},n)$. Now we repeat the bootstrap sampling $B$ times and the distribution of $\hat{\theta}^{*}_{\hat{\Psi}}(\textbf{X}^{*},n)$ is the bootstrap estimator of the distribution of $\hat{\theta}_{\Psi}(\textbf{X},n)$. With this estimated distribution we can also obtain the variance, the bias and the confidence intervals of our estimator. \\
Formally, the bootstrap methodology is based in the plug-in principle. The parameter of interest can be written as a function of the probability model, $\theta = f(\Psi)$. As the probability model is unknown, the plug-in estimate of our parameter is defined to be $\hat{\theta} = f(\hat{\Psi})$. So the bootstrap propose that we resample from this estimated probability model $\hat{\Psi}$ that is chosen to be close to $\Psi$ in some sense. 

If we have some information about $\Psi$ besides the data, the chosen $\hat{\Psi}$ must contain this information. Suppose that we know that the data $\textbf{X} = \{x_t\}_{t=1}^n$ comes from a certain process ruled by a probabilistic model $\Psi$ that depends on a finite number of parameters $\mathbf{\Xi} = \{\xi\}_{i=1}^k$, so $\Psi~=~f\left(\mathbf{\Xi}\right)$. This parameters can be estimated in the traditional statistical parametric approach as Maximum Likelihood getting $\mathbf{\hat{\Xi}} = \{\hat{\xi}\}_{i=1}^k$, and equivalently  $\hat{\Psi}~=~f\left(\mathbf{\hat{\xi}}\right)$. 

Now the bootstrap samples $\textbf{X}^{*} = \{x_1^{*},x_2^{*} \dots x_n^{*}\}$ comes from a process ruled by a probabilistic model $\hat{\Psi}$ (Fig. \ref{fig:bootscheme}).
These bootstrap samples emulates in every sense the original samples, including the correlation between the values.

\subsection{The transition probabilities of a symbol sequence}
\label{transition}

 As stated in Section \ref{sec:permu} using the methodology proposed by Bandt \& Pompe, the dynamics of a process $\{X_t\}_{t \in T}$ with $X_t \in \mathbb{R}$ is represented by a $m!-th$ finite state random process $\{S_t\}_{t \in (T-m+1)}$ with $ S_t \in S_m = \{\pi_1,\pi_2 \dots \pi_{m!}\}$ for all posible $m\geq 2$.
This realization of the symbolic sequence is thought to be produced by a probabilistic model with a fixed transition probability  $P^{ij}$ (i.e. the probability of moving from a symbol $\pi_i$ to a symbol $\pi_j$ for all $1 \leq i \leq m!$ and $1 \leq j \leq m!$) denoted $\Psi(P^{ij})$. 
 
We estimate the model $\hat{\Psi} = \Psi(\hat{P^{ij}})$, see Fig. \ref{fig:bootscheme} to bootstrap the Permutation Entropy

According to \citep{bandt2002permutation}, the relative frequency

\begin{equation}\label{eq:frequ}
\hat{P}_T(\pi_i)~=~
\frac{n_i}{T-m+1} 
\end{equation}
 
is an estimate \textit{as good as possible for a finite series of values}  of $P(\pi_i)$, with $n_i$ the number of times the state $\pi_i$ is observed up to time $T-m+1$. The sub-index $T$ in $\hat{P}_T(\pi_i)$ reinforces the notion of the dependence of the estimator on the length of the series $T$.

With the same spirit we define the transition probabilities of the symbol sequence as:

\begin{equation}
P^{ij}=P(s_{t+1}=\pi_j | s_t= \pi_i) ~~~ {1 \leq i \leq j \leq  m!}
\end{equation}

And the estimator of $P^{ij}$ 

\begin{equation}\label{eq:Pij}
\hat{P}_T^{ij} = 
\begin{cases*}
      \frac {n_{ij}}{n_i} ~~ if ~ n_i \geq 0\\
      0     ~~~     otherwise
    \end{cases*}
\end{equation}

where $n_{ij}$ is the number of transitions observed form $\pi_i$ to $\pi_j$ up to time $T-m+1$ and . Note $n_i~=~ \hat{P}(\pi_i).(T-m+1)$

Then by the law of the total probability:

\begin{equation}
P(\pi_j)~=~\sum_{i=1}^{m!} P(\pi_i) P^{ij}
\end{equation}

so if we call $\mathbf{P}(\pi)$ to the (m!)-dimensional vector containing $P(\pi_i)$ in each coordinate (i.e  $\mathbb{P}(\pi)~=(P(\pi_1),P(\pi_2),\dots,P(\pi_{m!})$), then $\mathbf{P}(\pi)$  is determined by $P^{ij}$, leading to the conclusion that the estimator $\hat{\mathbf{P}}_T(\pi)$ is determined by the estimation of $\hat{P}_T^{ij}$.

\subsection{Bootstrapping the Permutation Entropy}
\label{boot2}

The Permutation Entropy is defined in eq. \ref{shannon-disc}, so because of the plug-in principle, our natural estimator is:

\begin{equation}\label{eq:hatPE}
 \hat{\mathcal{H}}_T~=~\left\{-\sum_{i=1}^{N}~ \hat{P}_T(\pi_i) \ln(\hat{P}_T(\pi_i)) \right\} /  ln (m!)
\end{equation}

In section \ref{transition} we showed that the Permutation Entropy was completely defined by the transition probabilities $P^{ij}$ so we can think of them as  parameters of a probabilistic model $\Psi$. 

Following the scheme in Fig. \ref{fig:bootscheme} we have:\\

$\Psi \left(P^{ij}\right) \longrightarrow \mathbf{S}=(s_1,s_2\dots,s_{T-m+1}) \longrightarrow  \hat{\mathcal{H}}_T$\\

Our probabilistic model with unknown transition probabilities $P^{ij}$ gives the observed symbol sequence $\mathbf{S}$, and with that sequence the estimation of the Permutation Entropy is obtained.\\
 In the 'bootstrap world":\\

\begin{small}
 $\hat{\Psi} ~=~ \Psi \left(\hat{P}_T^{ij}\right) \longrightarrow \mathbf{S^{*}}=(s_1^{*},s_2^{*}\dots,s^{*}_{T-m+1}) \longrightarrow  \hat{\mathcal{H}}^{*}_T$\\
\end{small}

 $\hat{\Psi}$ generates $\mathbf{S^{*}}$ by a simulation, giving the bootstrap replication $ \hat{\mathcal{H}}^{*}_T$. We can repeat the simulation to get as many bootstrap replications as affordable.
  

Computing $B$ bootstrap replication of the permutation entropy from a time series $\{X_t\}_{t \in T}$ is simple: given a time series of lenght $T$, choose a world length $m$ and a time delay $\tau$ to do the mapping from $\{X_t\}_{t \in T}$ to $\{S_t\}_{t \in (T-m+1)}$ as stated in section \ref{sec:permu}. With this sequence: compute  $\hat{P}_T(\pi_i)$, (eq. \ref{eq:frequ}), $\hat{P}_T^{ij}$ (eq. \ref{eq:Pij}) and calculate  $\hat{\mathcal H}_T$ (eq. \ref{eq:hatPE}). Then choose at random with probability $\hat{P}_T(\pi)$ an inicial state $s^{*}_1(b)= \pi_k $ and choose at random with probability $\hat{P}_T^{kj}$ (note that $k$ is fixed with the value of the previous state) the next simulated state $s_2^{*}(b)$.
Repeat this last step $T-m+1$ times to obtain the simulation $\mathbf{S^{*}(1)}=(s_1^{*}(1),s_2^{*}(1)\dots,s^{*}_{T-m+1}(1))$. With this bootstrap replication of symbol sequence estimate $\hat{\mathcal H}^{*}_T(b)$ (equation \ref{eq:frequ}).

For a more detailed reference see Algorithm \ref{alg:algoritmoRaro} in Appendix \ref{Appen}.

Repeat the simulation of the sequence $B$ times to obtain $\hat{H}^{*}_T(b) ~~ b =  1 \dots B$. With the set $\hat{\mathcal H}^{*}_T(b) ~~ b =  1 \dots B$ we have the bootstrap replications needed to estimate the standard deviation, the confidence intervals of  $\hat{\mathcal H}_T$, or the test presented in the following section.

So we obtained $B$ bootstrap replications of $\hat{\mathcal{H}}^{*}_T$: 

$\hat{\mathcal{H}}^{*}_T(1),\hat{\mathcal{H}}^{*}_T(2)\dots\hat{\mathcal{H}}^{*}_T(B)$

The Bootstrap Standard Deviation of $\hat{\mathcal{H}}^{*}_T$ is our estimation of the Standard Deviation of $\hat{\mathcal{H}}_T$:
\begin{equation}
\hat{\sigma}_B(\hat{\mathcal{H}}_T)~=~\hat{\sigma}(\hat{\mathcal{H}}^{*}_T)
\end{equation}
and is defined as
\begin{equation}
\label{eq:bootsd}
\hat{\sigma}(\hat{\mathcal{H}}^{*}_T)=\sqrt{\frac{1}{B-1}\sum_{i=1}^{B} \left(\hat{\mathcal{H}}_T^{*}(i) - \hat{\mathcal{H}}^{*}_T(\bullet)\right)^2}
\end{equation}
where
\begin{equation}
\label{eq:bootmean}
\hat{\mathcal{H}}^{*}_T(\bullet)~=~\frac{1}{B}\sum_{i=1}^{B}\hat{\mathcal{H}}^{*}_T(i)
\end{equation}

We define the bootstrap bias of $\hat{\mathcal{H}}^{*}_T$ as:

\begin{equation}
\label{eq:bias}
\text{Bias}(\hat{\mathcal{H}}^{*}_T)=\hat{\mathcal{H}}^{*}_T(\bullet) - \hat{\mathcal{H}}_T
\end{equation}

Finally, the Mean Square Error (MSE) of an estimator:
\begin{equation}
\text{MSE}(\hat{\mathcal{H}}^{*}_T)= Var(\hat{\mathcal{H}}^{*}_T)+Bias^2(\hat{\mathcal{H}}^{*}_T)
\end{equation}

\subsubsection{Confidence Intervals}
The $1-\alpha$ Confidence Interval of $\mathcal{H}$ is defined by the percentiles of the bootstrap $\delta$. For each bootstrap replicate $\hat{\mathcal{H}}^{*}_T(b)$ we compute the difference - $\delta^{*}(b)$ - between that replication and the mean of all bootstrap replicates. 
Then we choose the ($\frac{\alpha}{2}$) and the ($1-\frac{\alpha}{2}$) percentiles of the $\delta^{*}$'s distribution and add them to the original estimate, $\hat{\mathcal{H}}^{*}_T$, correcting for the bias, and the resulting $(1-\alpha)100\%$ confidence interval is:

\begin{equation}
\left[\max(2 . \hat{\mathcal{H}}_T - \hat{\mathcal{H}}^{*}_T(\bullet) +\delta^{*}_{\frac{\alpha}{2}},0),\min(2 . \hat{\mathcal{H}}_T - \hat{\mathcal{H}}^{*}_T(\bullet) +\delta^{*}_{(1-\frac{\alpha}{2})},1)\right]
\end{equation}\label{eq:bootci}

For a more detailed reference see Algorithm \ref{alg:algoritmo2} in Appendix \ref{Appen}.

\subsubsection{Hypothesis testing}

With this same spirit, a confidence interval for the difference between the permutation entropy of two different time series can be made. In inferential statistics exists a direct relationship between confidence intervals and hypothesis testing. A two-sided $(1-\alpha)$ confidence interval in the difference between two measures can be used to determine if those two measures are significantly different by only checking if the $zero$ belongs to this particular interval.
\\
\\
$H_0:\Delta = \mathcal{H}_1 - \mathcal{H}_2 = 0$
\\
\\
If $0~ \notin~ (1-\alpha)100\% ~CI~(\Delta)$ \\then reject $H_0$ and $$\mathcal{H}_1 \neq \mathcal{H}_2$$

The procedure to perform this test is shown in \ref{alg:hypothesis} in Appendix \ref{Appen}.

\section{Numerical simulation}
\label{numerical}
In order to show our proposed bootstrap in a very general time series, we simulate a well known dynamical system: the $ 1/f^{\alpha}$. All the series are simulated with different time spam, $T$ in order to evaluate the statistical properties of  $\hat{\mathcal{H}}_T$ according to Ec. \ref{eq:hatPE}.
As stated before, a way to obtain the distribution of $\hat{\mathcal{H}}$ is to repeat the experiment a large number of times and approximate the distribution of $\hat{\mathcal{H}}$ by the so obtained empirical distribution. While for real world experiments this can be inapplicable, for simulated time series this can easily done by Montecarlo Simulation. 
Once the $n$ replications of $\hat{\mathcal{H}_T}~=~\{\hat{\mathcal{H}_T}(1),\dots,\hat{\mathcal{H}_T}(n)\}$ is obtained the standard deviation is estimated by:
\begin{equation}
\label{eq:sd}
\hat{\sigma}(\hat{\mathcal{H}}_T)=\sqrt{\frac{1}{n-1} \sum_{i=1}^{n} \left(\hat{\mathcal H}_T(i)-\hat{\mathcal H}_T(\bullet\right)^2}
\end{equation}
where
\begin{equation}
\label{eq:mean}
\hat{\mathcal{H}}_T(\bullet)~=~\frac{1}{n}\sum_{i=1}^{n}\hat{\mathcal{H}}_T(i)
\end{equation}

\subsection{Experimental design}

\begin{figure}
\includegraphics[width=\textwidth]{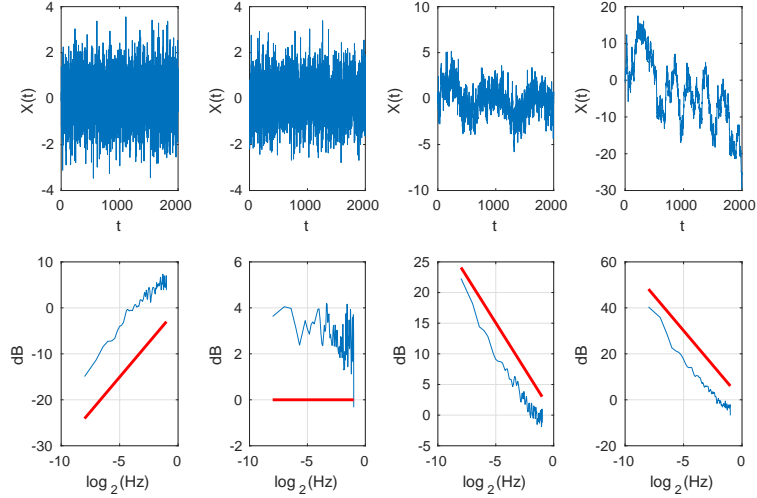}
\caption{(top) realization for $1/f^k$ noises (T=2000), from left to right: $k=-1$,$k=0$,$k=1$,$k=2$. (bottom) spectral density of the respective $1/f^k$ noises.} 
\label{fig:ruidos}
\end{figure}

\paragraph{Stochastic dynamical systems:}\label{stochasticdesign}$1/f^{\alpha}$ noises refers to a signal with spectral density $S(f)$ with the form $S(f)~=~k \frac{1}{f^{\alpha}}$ where $k$ is a constant, $\alpha$ is the signal-dependent parameter and $f$ is frecuency \citep{kasdin1995}. It is a stochastic model which seems to be ubiquitous in nature \citep{kasdin1995} and the references therein. We simulate $1/f^{\alpha}$ noises with $\alpha=\{-1,0,1,2\}$. See Fig. \ref{fig:ruidos} for an example of these noises. A \textit{white noise} process ($\alpha = 0$) would generate a curve with constant power in the spectrum. The case of $\alpha = 1$ or \textit{pink noise} is the canonical case and of most interest as many of the values of $\alpha$ found in nature are very near to 1.0 \citep{caloyannides1974,musa1981,novikov97,vossclarke75,dutta1981}. A random walk noise (Brownian Motion or \textit{red noise}, $\alpha=2$) would show a $(1/f^2)$ distribution in $S(f)$. In order to simulate this stochastic process, the algorithm propose in \citep{timmer1995generating} is used.

%

For each $\alpha=\{-1,0,1,2\}$ 1000 replications were simulated for each 
$T=\{60,100,120,400,$ $600,2000,3600,5000,10000,20000,50000\}$.\\
 $\hat{\mathcal{H}}_i(T,m)$ for $\{i=1\dots 1000\}$ along with $\hat{\sigma}(\hat{\mathcal{H}}_T)$ are obtained for $m=\{3,4,5,6\}$.\\
 For each $\alpha=\{-1,0,1,2\}$ a single replication was simulated for each $T=\{60,100,120,400,600,2000,3600,5000,$ $10000,20000,50000\}$. In each case for this replication we implemented the algorithm \ref{alg:algoritmoRaro} to get 1000 bootstrap replicates. $\hat{\mathcal{H}}_i^*(T,m)$ for $\{b=1 \dots 1000\}$ and $\hat{\sigma}_B(\hat{\mathcal{H}}_T)$ are obtained for $m=\{3,4,5,6\}$.
For these bootstrap distribution we analyze \textit{Bias}, \textit{Standard Deviation} and \textit{MSE}. 
 \\
As for each set of 1000 bootstrap replicates we obtain a single confidence interval, we repeated this step 50 times to obtain Table \ref{tab:IC} that indicates the estimated confidence level of this method along with the mean amplitude of the interval. 

\section{Application: EEG data}\label{EEGdata}

In order to illustrate the proposed confidence intervals in real contexts we present how it can
describe the variability in the Permutation Entropy within one observation of Electroencephalogram (EEG) Data. More precisely, as a first practical application, we analyze, via PME, four different sets of EEGs for healthy and epileptic patients that were
previously analyzed by \citep{andrzejak2001indications}\\ (available at
http://www.meb.unibonn.de/epileptologie/science/physik/eegdata.html). The data consist of 100 data segments (from which we choose 10 at random), whose length is 4097 data points with a sampling frequency of 173.61Hz, of brain activity for different groups and recording regions: surface EEG recordings from five healthy volunteers in an awake state with eyes open (Set A) and closed (Set B), intracranial EEG recordings from five epilepsy patients during the seizure free interval from outside (Set C) and from within (Set D) the seizure generating area. Details about the recording technique of these EEG data can be found in the original paper.

\section{Results and discussion}
\label{results}


We intend to show in our simulated experiment that the bootstrap distribution of the PE estimator is close in every meaningful sense to the distribution obtaining by the repetition of the original experiment (empirical distribution) in order to obtain this estimator distribution when the exact replication of the experiment can not be done.
 A comparison between the standard deviations of both bootstrap replicates and the empirical distribution ($\hat{\sigma}_B(\hat{\mathcal{H}}_T)$ and $\hat{\sigma}(\hat{\mathcal{H}}_T)$ respectively) for the stochastic processes is presented in Fig. \ref{fig:vardividido}. There are some discrepancies for low values of $T$, but form a certain value $T_0$ in all the cases of $m$ the standard deviation coincides. In Fig. \ref{fig:sesgo} it can be seen that for every $m$ and $\alpha$ the bias of the bootstrap estimate tends to zero as $T$ increases. So, this bootstrap estimator is an asymptotically unbiased estimator. With this and with the fact that $\sigma$ also goes to zero as $T$ increases, the bootstrap estimator seems to be Mean Square Consistent. Even more, for large values of $T$, the bootstrap estimation is as efficient as the estimation produced by the repetition of the experiment.
 
In Fig. {\ref{fig:histo} and for an arbitrary value of $\alpha=1$ an for the largest length of the simulated time series $T=50000$, an histogram of the bootstrap estimator along with an histogram of the simulated estimator are presented in different scale for every $m$. The similar shape between the histograms can be appreciated, the difference in the location is due that the bootstrap samples depends on only one of the estimations of the PE (that are random) but this does not affect the posterior inferential conclusions.

 For a more thorough exploration of the bootstrap estimator we have calculated fifty $90\%$ Confidence Intervals and for every $m$ and every {$\alpha$} we computed how many times the real value of ${\mathcal{H}}$ - in fact we use the mean of $\hat{\mathcal{H}}(T,m)$ (see paragraph \ref{stochasticdesign}) that is the best possible estimator - is outside the bounds of the confidence interval.  Results are shown in Table \ref{tab:IC}. For white noise  the confidence level is in fact higher than 90\%, in fact is always accurate but for other values of $\alpha$ the overall confidence level is approximately between 90\%. and 95\%.

In many practical situations, there is a wish to compare the dynamics of two processes via the Permutation Entropy of their time series. The question is:  $\mathcal{H}_1 = \mathcal{H}_2$? This can not be answered with punctual estimators ($\hat{\mathcal{H}}_1 , \hat{\mathcal{H}}_2$) because these are continuous random variables and with probability 1 (i.e. \textit{always}) they are going to be different. The real question is if that difference is statistically significant or not, and that only can be answered if exists a measure of variability of that continuous random variable, $\hat{\Delta}=\hat{\mathcal{H}}_1 - \hat{\mathcal{H}}_2$. There has not been, up to our best knowledge, this kind of variability measure that we are proposing now. 
An example of this is the Permutation Entropy of EEG signals. 

The problem of interest is comparing the PE of 4 different sets of EEG signals: EEG signals of patients in an awake state with eyes open (Set A) and closed (Set B), intracranial EEG recordings from epilepsy patients during the seizure free interval from outside (Set C) and from within (Set D) the seizure generating area.

If many EEG signals for each type of patients can be recorded a classical inference for the mean can be performed if the normality assumptions are complied or, if normality fails (that is to be expected in this case), a non parametric test can be made. But is the mean PE representative of the population of each type of patient? 

A different problem is to analyze the variability of a single EEG signal, this can not be done with conventional methods and up to this contribution there has not been an answer to this problem. The same problematic applies when the issue is to compare between two EEG signals.

We solve this problem by constructing confidence intervals and hypothesis testing with our proposed method. In Fig \ref{fig:intervalosEEG} 90\% Confidence Intervals for the 10 EEG signals of brain activity for different groups and recording regions are performed. It should be pointed out that the overlapping between intervals does not necessarily means that there is no significant differences between the two Permutation Entropies. To reach that conclusion, an hypothesis test for the difference must be made.

In Fig. \ref{fig:testABEEG} we perform a test for difference in the Permutation Entropy between the 10 EEG signals of healthy volunteers in an awake state with eyes open ($Set A$) and the 10 EEG signals of healthy volunteers in an awake state with eyes closed ($Set B$). Each EEG signal of $Set A$ was compared with each signal of $Set B$ with a 10\% significance level, and the conclusion is that the differences seems to be at random, indicating that is no real difference between these two types of EEG signals. 
In Fig. \ref{fig:test} the same analysis is extended to all the different types of patients. While the differences between $Set A$ and $Set B$ seems to be at random, all EEG signals of those Sets are different in every test to the EEG signals of $Set C$ and $Set D$. Instead, between $Set C$ and $Set D$ again the differences are distributed between significant and not significant.
 
In summary, we present a computer based methodology to obtain an accuracy measure for the estimation of the permutation entropy ($\hat{\mathcal{H}}$). So far we found in the literature that only descriptive statistics are used to characterize this quantifier and if the objective is to extrapolate on and reach conclusions that extend beyond the raw data itself there were no statistical inference method at hand. Even a simple comparison between to random variables (as $\hat{\mathcal{H}}$) can not be made with some confidence without a measure of variability of that variable. Our method paves the way to perform any inferential statistic involving the Permutation Entropy or even any entropy that uses the Probability Function Distribution proposed by Bandt and Pompe.

\begin{table}
\caption{$90\%$ Confidence Intervals (Eq. \ref{eq:bootci}) for every symbol length $m$ and power law parameter $\alpha$. How many times the real value of ${\mathcal{H}}$  - in fact we use the mean of $\hat{\mathcal{H}}(T,m)$ (see paragraph \ref{stochasticdesign}) that is the best possible estimator - is lower than the Lower Bound $Miss Left$ or higher than the Upper Bound $Miss Right$. For white noise  the confidence level is in fact higher than 90\%, in fact is always accurate but for other values of $\alpha$ the overall confidence level is approximately between 90\%. and 95\%. }
\label{tab:IC}       
\begin{center}
\begin{tabular}{cccccc}
\hline\noalign{\smallskip}
m & $\alpha$ & $\hat{\mathcal{H}}(T,m)$ &  Miss left & MissRight & Mean Amplitude \\
\noalign{\smallskip}\hline\noalign{\smallskip}
3 & -1    & 0.995831848 & 0       & 0.04           & 0.00222              \\
4 & -1    & 0.989083439 & 0.02    & 0.04           & 0.00420              \\
5 & -1    & 0.983495069 & 0.04    & 0.04           & 0.00500              \\
6 & -1    & 0.97547007  & 0.02    & 0              & 0.00555              \\
3 & 0     & 0.99990292  & 0       & 0              & 0.00057              \\
4 & 0     & 0.999679839 & 0       & 0              & 0.00080              \\
5 & 0     & 0.998800463 & 0       & 0              & 0.00134              \\
6 & 0     & 0.994503528 & 0       & 0              & 0.00235              \\
3 & 1     & 0.991622896 & 0.02    & 0.02           & 0.00340              \\
4 & 1     & 0.983385433 & 0.02    & 0.02           & 0.00493              \\
5 & 1     & 0.97600538  & 0.02    & 0.04           & 0.00591              \\
6 & 1     & 0.966355927 & 0       & 0.06           & 0.00657              \\
3 & 2     & 0.943233315 & 0.08    & 0.06           & 0.00959              \\
4 & 2     & 0.90634703  & 0.04    & 0.02           & 0.01273              \\
5 & 2     & 0.878452628 & 0.02    & 0.02           & 0.01413              \\
6 & 2     & 0.853327039 & 0.04    & 0.02           & 0.01482             \\
\noalign{\smallskip}\hline
\end{tabular}
\end{center}
\end{table}

\begin{figure}[H]
\includegraphics[width=\textwidth]{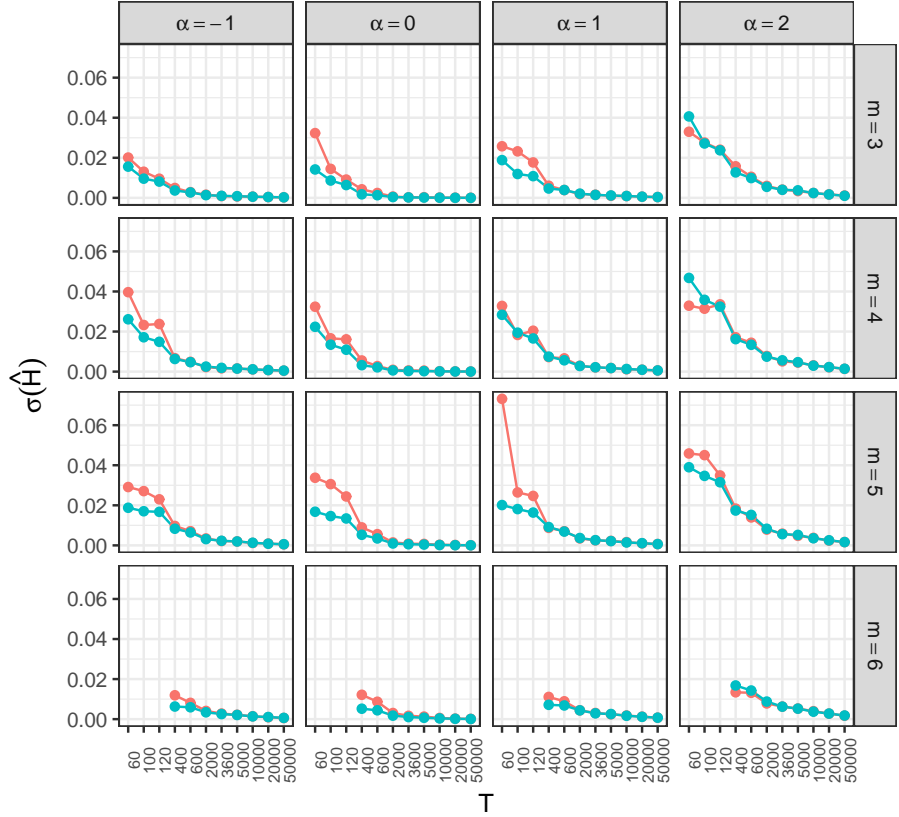}
\caption{Standard Deviation of $\hat{\mathcal{H}}$ as $T$ increases. A comparison between the standard deviations of $\hat{\mathcal{H}}$ of both bootstrap replicates in red and simulated replicates in blue is shown. There are some discrepancies for low values of $T$, but form a certain value $T_0$ in all the cases of symbol length $m$ and $\alpha$ the standard deviation coincide. As the bias goes to zero (Fig.\ref{fig:sesgo}) along with the standard deviation the bootstrap estimator seems to be a mean square consistent estimator.}
\label{fig:vardividido}
\end{figure}

 \begin{figure}[H]
\includegraphics[width=\textwidth]{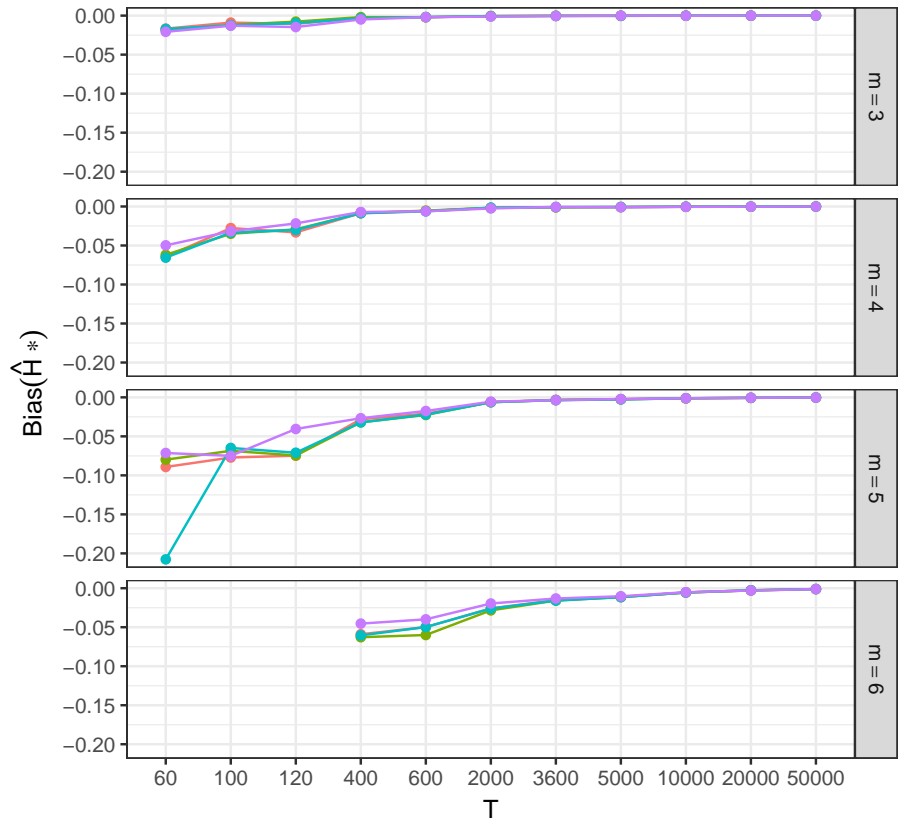}
\caption{The Bootstrap Bias for different values of $\alpha$ in function of $T$. It can be seen that for every symbol length $m$ and $\alpha$ the bias of the bootstrap estimate tends to zero as $T$ increases. The bootstrap estimator is an asymptotically unbiased estimator. }
\label{fig:sesgo}
\end{figure}

\begin{figure}[H]
\includegraphics[width=\textwidth]{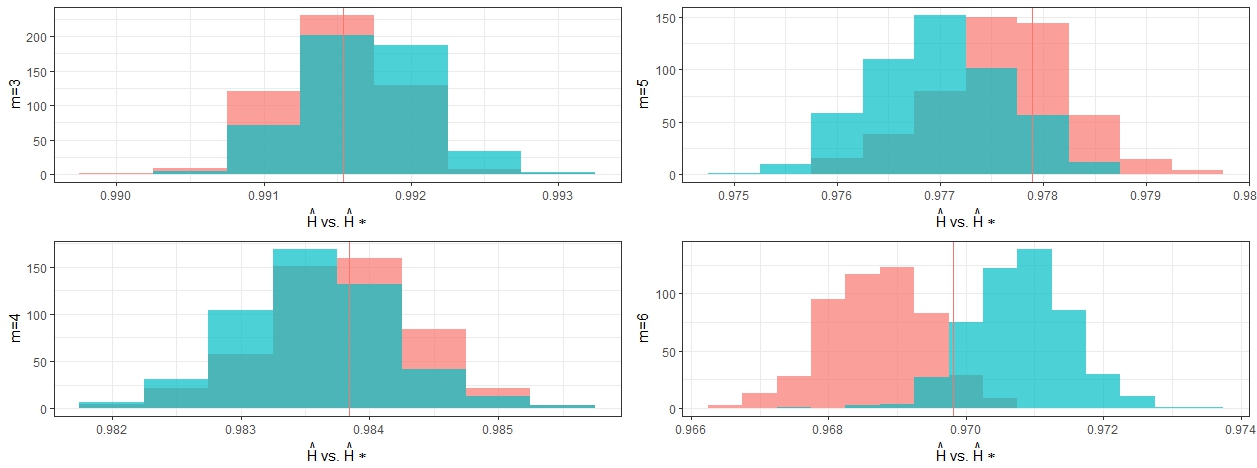}
\caption{For an arbitrary value of $\alpha=1$ an for the largest length of the simulated time series $T=50000$, an histogram of the bootstrap estimator (Red) along with an histogram of the simulated estimator (Blue) are presented in different scales for every $m$. The similar shape between the histograms can be appreciated, the difference in the location is due that the bootstrap samples depends on only one of the estimations of the PE (that are random) but this does not affect the posterior inferential conclusions. }
\label{fig:histo}
\end{figure}

\begin{figure}[H]
\includegraphics[width=\textwidth]{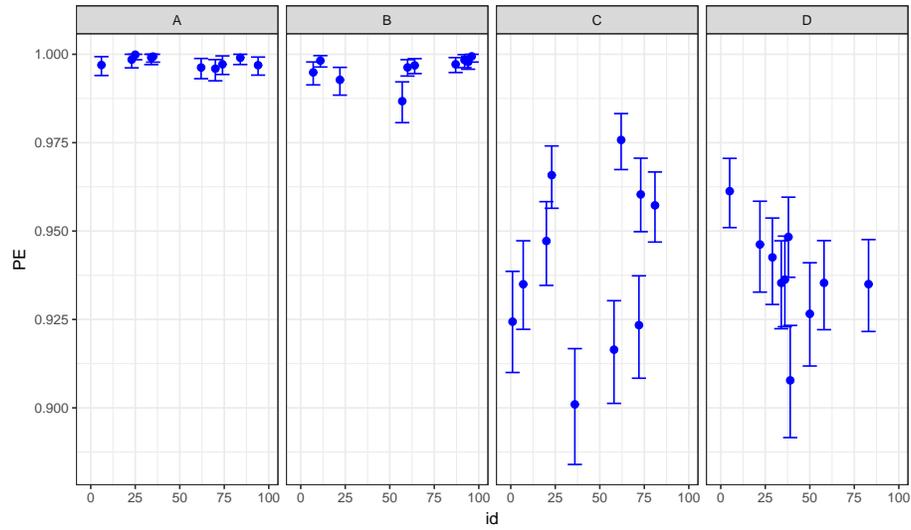}
\caption{The 90\% Confidence Intervals for the 10 EEG signals of brain activity for different groups and recording regions: surface EEG recordings from healthy volunteers in an awake state with eyes open (Set A) and closed (Set B), intracranial EEG recordings from epilepsy patients during the seizure free interval from outside (Set C) and from within (Set D) the seizure generating area. It should be pointed out that the overlapping between intervals does not necessarily means that there is no significant differences between the two Permutation Entropies. To reach that conclusion, an hypothesis test for the difference must be made. }
\label{fig:intervalosEEG}
\end{figure}

\begin{figure}[H]
\includegraphics[width=\textwidth]{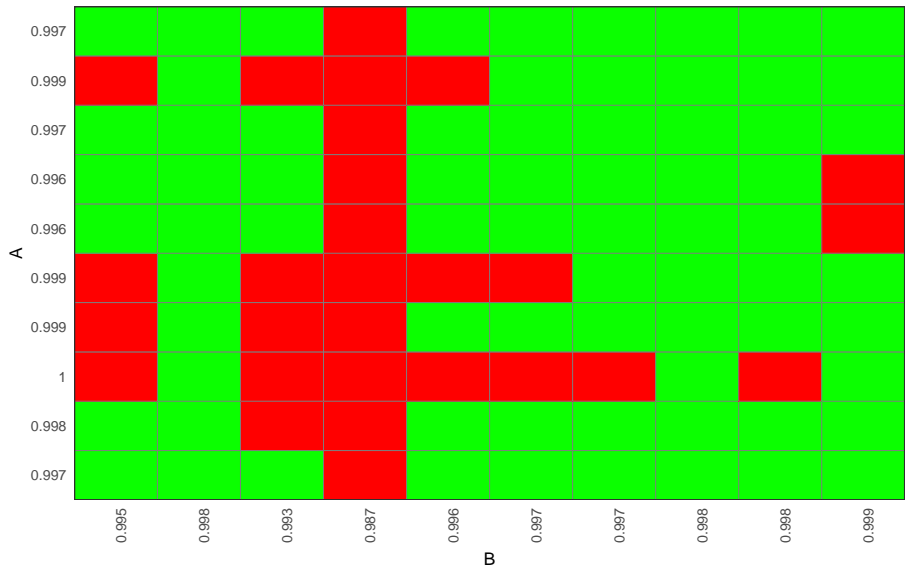}
\caption{Hypothesis Test: Difference in the Permutation Entropy of a time series. A test for difference in the Permutation Entropy between the 10 EEG signals of healthy volunteers in an awake state with eyes open ($Set A$) and the 10 EEG signals of healthy volunteers in an awake state with eyes closed ($Set B$) is shown at the top left. Each EEG signal of $Set A$ was compared with each signal of $Set B$ with a 10\% significance level, and the results are shown. The red squares mean that the test was rejected and there is a significant difference between the Permutation Entropies. On the other side green squares mean that the test was not rejected and there is no evidence for that difference.
It should be pointed out that this is not a test for the difference in the mean Permutation Entropy of all EEG signals in $Set A$ vs all EEG signals in $Set B$, but instead a $one-on-one$ test between the Permutation Entropy for each single EEG signal of $Set A$ vs. the Permutation Entropy for each single EEG signal of $Set B$ repeated, giving a total of 100 tests. In the $x-axis$ are the estimations of the Permutation Entropy of each signal of the $Set B$ EEG signals, and on the $y-axis$ are the estimations of the Permutation Entropy of each signal of the $Set a$ EEG signals.}
\label{fig:testABEEG}
\end{figure}

\begin{figure}[H]
\includegraphics[width=\textwidth]{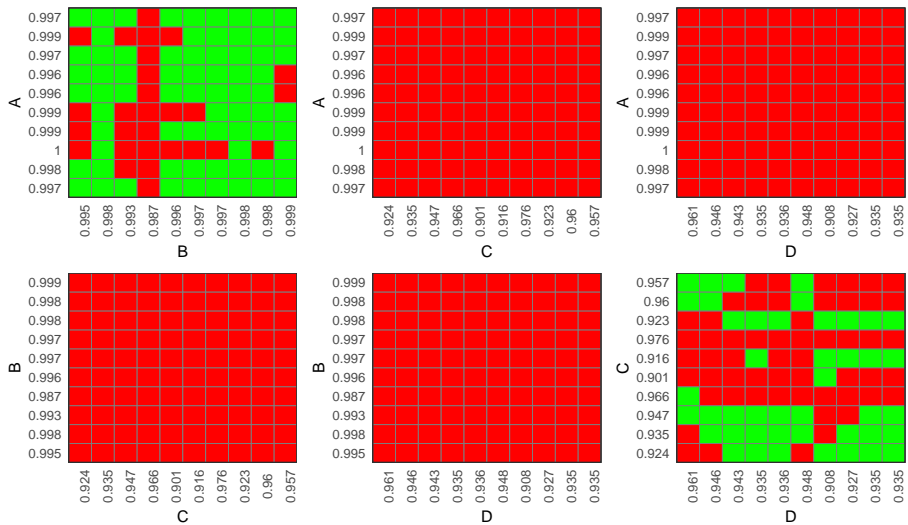}
\caption{The same analysis of the previous figure is extended to all the different types of patients. While the differences between $Set A$ and $Set B$ seems to be at random, all EEG signals of those Sets are different in every test to the EEG signals of $Set C$ and $Set D$. Instead, between $Set C$ and $Set D$ again the differences are distributed between significant and not significant.}
\label{fig:test}
\end{figure}


\section*{References}



\newpage
\appendix
\section{Algorithms}\label{Appen}

\begin{algorithm}
\begin{algorithmic}[1]
\STATE $T \leftarrow $ time series length 
\STATE \textbf{set} $m$
\STATE \textbf{set} $\tau$
\STATE \textbf{compute} $\hat{P}_T(\pi_i)$ (Eq. \ref{eq:frequ}) from the actual time series 
\STATE  \textbf{compute} $\hat{\mathcal H}_T$ (Eq. \ref{eq:hatPE}) from the actual time series
\STATE \textbf{compute} $\hat{P}_T^{ij}$ (Eq. \ref{eq:Pij}) from the actual time series
\STATE $b \leftarrow 1$
\WHILE {$b \leq B$}
\STATE $i \leftarrow 1$
\STATE $s_i^{*}(b) \leftarrow \pi_k $ w.p. $\hat{P}_T(\bm{\pi})$
\COMMENT {i. e. the initial state for the b$-th$ bootstrap replication}
\WHILE {$i \leq T-m+1$}
\STATE $ s_{(i+1)}^{*}(b) \leftarrow \pi_k $ w.p. $\hat{P}_T^{ik}(\bm{\pi})$
\COMMENT {i. e. the i$-th$ state for the b$-th$ bootstrap replication}
\STATE $i \leftarrow i+1$
\ENDWHILE
\STATE \textbf{estimate} $\hat{P}^{*}( \bm{\pi})$ using $\mathbf{S}^{*}(b)$ $ $ Ec. 7 
\STATE \textbf{estimate} $\hat{\mathcal H}^{*}_T(b)$ using $\hat{P}^{*}( \bm{\pi})$ and Ec. 11
\COMMENT {i. e. the b bootstrap sample of $\hat{\mathcal H}_T$.}
\ENDWHILE
\end{algorithmic}
\caption{Algorithm for the parametric boostrap for Permutation Entropy}\label{alg:algoritmoRaro}
\end{algorithm}
\begin{algorithm}
\begin{algorithmic}[1]
\WHILE {$b \leq B$}
\STATE \textbf{generate} $\hat{\mathcal{H}}^{*}_T(b)$\\
\ENDWHILE
\STATE \textbf{compute} $\hat{\mathcal{H}}^{*}_T(\bullet)~=~\frac{1}{B}\sum_{i=1}^{B}\hat{\mathcal{H}}^{*}_T(i)$
\STATE \textbf{sort} $\delta^{*}(b)=\hat{\mathcal{H}}^{*}_T(b)-\hat{\mathcal{H}}^{*}_T(\bullet)$ in increasing order
\STATE \textbf{set} confidence level $1-\alpha$
\STATE \textbf{compute} $\delta^{*}_{ \frac{\alpha}{2} } \leftarrow \left\{ \delta^{*}_{ \frac{\alpha}{2} } \bigg/ \frac{\#~\left(\delta^{*}<\delta_{\frac{\alpha}{2}}\right)}{B} \leq \frac{\alpha}{2} \right \}$\\
\COMMENT {i. e. if $B=1000$ and $\alpha = 0.1$ choose the $50th$ element on the sorted $\delta^{*}$ }
\STATE \textbf{compute}\\ $\delta^{*}_{ (1-\frac{\alpha}{2}) } \leftarrow \left\{\delta^{*}_{ (1-\frac{\alpha}{2})}  \bigg/ \frac{\#~\left(\delta^{*}<\delta_{(1-\frac{\alpha}{2})}\right)}{B} \leq 1-\frac{\alpha}{2} \right \}$\\
\COMMENT {i. e. if $B=1000$ and $\alpha = 0.1$ choose the $950th$ element on the sorted $\delta^{*}$ }
\STATE The lower bound of the confidence interval is \\$\max(2.\hat{\mathcal{H}}_T-\hat{\mathcal{H}}^{*}_T(\bullet)+\delta^{*}_{\frac{\alpha}{2}},0)$
\STATE The upper bound of the confidence interval is \\$\min(2.\hat{\mathcal{H}}_T-\hat{\mathcal{H}}^{*}_T(\bullet)+\delta^{*}_{(1-\frac{\alpha}{2})},1)$
\end{algorithmic}
\caption{Algorithm for the confidence interval for Permutation Entropy}\label{alg:algoritmo2}
\end{algorithm}

\begin{algorithm}
\begin{algorithmic}[1]
\STATE \textbf{compute} $\hat{\mathcal{H}_1}_T$ the PE of the $1st$ time series
\STATE \textbf{compute} $\hat{\mathcal{H}_2}_T$ the PE of the $2nd$ time series
\STATE \textbf{compute} $\hat{\Delta_T}=\hat{\mathcal{H}_1}_T-\hat{\mathcal{H}_2}_T$
\WHILE {$b \leq B$}
\STATE \textbf{generate} $\hat{\mathcal{H}_1}^{*}_T(b)$ the bootstrap replicate of the $1st$ time series
\STATE \textbf{generate} $\hat{\mathcal{H}_2}^{*}_T(b)$ the bootstrap replicate of the $2nd$ time series
\ENDWHILE
\FOR{i in 1 to B}
      \FOR{k in 1 to B}
        \STATE \textbf{compute} $\Delta_T^{*}(n) = \hat{\mathcal{H}_1}^{*}_T(i)-\hat{\mathcal{H}_2}^{*}_T(k)$
      \ENDFOR
    \ENDFOR
\STATE \textbf{compute} $\hat{\Delta}^{*}_T(\bullet)~=~\frac{1}{B^2}\sum_{i=1}^{B^2}\hat{\Delta}^{*}_T(n)$
\STATE \textbf{sort} $\delta^{*}(n)=\Delta_T^{*}(n)-\hat{\Delta}^{*}_T(\bullet)$ in increasing order
\STATE \textbf{set} confidence level $1-\alpha$
\STATE \textbf{compute} $\delta^{*}_{ \frac{\alpha}{2} } \leftarrow \left\{ \delta^{*}_{ \frac{\alpha}{2} } \bigg/ \frac{\#~\left(\delta^{*}<\delta_{\frac{\alpha}{2}}\right)}{B} \leq \frac{\alpha}{2} \right \}$\\
\COMMENT {i. e. if $B=1000$ and $\alpha = 0.1$ choose the $50th$ element on the sorted $\delta^{*}$ }
\STATE \textbf{compute}\\ $\delta^{*}_{ (1-\frac{\alpha}{2}) } \leftarrow \left\{\delta^{*}_{ (1-\frac{\alpha}{2})}  \bigg/ \frac{\#~\left(\delta^{*}<\delta_{(1-\frac{\alpha}{2})}\right)}{B} \leq 1-\frac{\alpha}{2} \right \}$\\
\COMMENT {i. e. if $B=1000$ and $\alpha = 0.1$ choose the $950th$ element on the sorted $\delta^{*}$ }

\STATE The lower bound of the confidence interval is \\$\hat{\Delta}_T+\delta^{*}_{\frac{\alpha}{2}}$
\STATE The upper bound of the confidence interval is \\$\hat{\Delta}_T+\delta^{*}_{(1-\frac{\alpha}{2}})$
\STATE If $0$ does not belong to the interval\\ Then $\mathcal{H}_1 \neq \mathcal{H}_2$ with $\alpha$ level of signification.
\end{algorithmic}
\caption{Algorithm for the hypothesis testing for Permutation Entropy}\label{alg:hypothesis}
\end{algorithm}

\end{document}